\documentclass[a4paper,12pt]{article}
\pdfoutput=1 
\usepackage{graphicx}
\usepackage{jheppub}
\usepackage{ytableau}
\usepackage{pgf,pgfarrows,pgfnodes}
\usepackage{graphics}
\usepackage{hyperref}
\usepackage{tikz}
\usepackage{xparse}
\usepackage{subfigure}
\newcommand{\be}{\begin{equation}}
\newcommand{\eeq}{\end{equation}}
\newcommand{\bea}{\begin{eqnarray}}
\newcommand{\eea}{\end{eqnarray}}
\newcommand{\ba}{\begin{array}}
\newcommand{\ea}{\end{array}}

\newcommand{\ee}{\end{equation} }

\newcommand{\tr}{\mathrm{tr}\,}

\newcommand{\one}{{\rm 1\kern -.9mm l}}

\newcommand{\CommaBin}{\mathbin{\raisebox{0.5ex}{,}}}

\title{Recurrence relations for the ${\cal W}_3$ conformal blocks
and ${\cal N}=2$ SYM partition functions}

\author{Rubik Poghossian}

\affiliation{Yerevan Physics Institute,\\
Alikhanian Br. 2, AM-0036 Yerevan, Armenia}
\emailAdd{poghos@yerphi.am}

\abstract{Recursion relations for the sphere $4$-point and torus
$1$-point ${\cal W}_3$ conformal
blocks, generalizing Alexei Zamolodchikov's famous relation
for the Virasoro conformal blocks are proposed.
One of these relations
is valid for any 4-point conformal block with two arbitrary
and two special primaries with charge parameters proportional
to the highest weight of the fundamental irrep of $SU(3)$.
The other relation is designed for the torus conformal block
with a special (in above mentioned sense) primary field insertion.
AGT relation maps the sphere
conformal block and the torus block to the instanton partition
functions of the ${\cal N}=2$ $SU(3)$ SYM theory with 6 fundamental
or an adjoint hypermultiplets respectively.
AGT duality
played a central role in establishing these recurrence relations,
whose gauge theory counterparts are novel
relations for the $SU(3)$ partition functions with $N_f=6$
fundamental or an adjoint hypermultiplets. By decoupling some (or all)
hypermultiplets, recurrence relations for the asymptotically free
theories with $0\le N_f<6$ are found.
}


\preprint{YerPhI/2017/04}
\dedicated{To the memory of Alexei Zamolodchikov}

\begin{document}

\maketitle
\flushbottom
\section{Introduction}
Conformal blocks play central role in any 2d CFT since they are
holomorphic building constituents of the correlation functions
of primary fields \cite{Belavin:1984vu}. In the case when the theory
possesses no extra holomorphic current besides the spin $2$
energy-momentum tensor, the conformal block is fixed by
the Virasoro symmetry solely. However a direct computation is
practical up to first few levels of the intermediate state.
Upon increasing the level such computation soon becomes intractable.
Some three decades ago Alexei Zamolodchikov found a brilliant solution
to this problem. Based on analysis of the poles and respective
residues of the $4$-point conformal block considered as a function
of the intermediate conformal weight and thorough investigation
of the semiclassical limit, a very efficient recursion formula has been
discovered \cite{Zamolodchikov:1985ie,Zamolodchikov:1987tmf}.
Successful applications of this recurrence relation include
Liouville theory \cite{Zamolodchikov:1995aa}, 4d ${\cal N}=2$ SYM
\cite{Poghossian:2009mk},
topological strings \cite{KashaniPoor:2012wb,Kashani-Poor:2013oza},
partition function and
Donaldson polynomials on $\mathbb{CP}_2$ \cite{Bershtein:2016mxz} et al.

Analogous recurrence relations has been found much later also for
torus $1$-point Virasoro block \cite{Poghossian:2009mk}
(see also \cite{Hadasz:2009db}) and for ${\cal N}=1$
Super-conformal blocks \cite{Belavin:2006zr,Hadasz:2007nt}.

The case when the theory admits higher spin ${\cal W}$-algebra symmetry
\cite{Zamolodchikov:1985wn,Fateev:1987zh,Fateev:2007ab}
is much more complicated. Holomorphic blocks of
correlation functions of generic ${\cal W}$-primary fields can not be
found on the basis of the ${\cal W}$-algebra Ward identities solely. Still,
it is known that if an $n$-point ($n\ge 4$) contains $n-2$
partially degenerate primaries\footnote{
In this paper the term partially degenerate refers to
the primary fields which admit a single null-vector on
level $1$.}, the ${\cal W}$-algebra is restrictive enough
to determine (in principle) such blocks. It appears that exactly
at this situation an alternative way to obtain ${\cal W}$-conformal blocks
based on AGT relation \cite{Alday:2009aq,Wyllard:2009hg,Fateev:2011hq}
is available.

Note that though AGT relations provide combinatorial formulae
for computing such conformal blocks, a recursion formulae like the one
originally proposed by Zamolodchikov have an obvious advantage.
Besides being very efficient for numerical calculations
\cite{Zamolodchikov:1995aa},
such recursive formulae are very well suited for the investigation
of analyticity properties and asymptotic behavior of the conformal
blocks (or their AGT dual instanton partition functions
\cite{Poghossian:2009mk}).
Instead the individual terms of the instanton sum have many spurious
poles that cancel out only after summing over all, rapidly
growing number of terms of given order which leaves the final analytic
structure more obscure.

In this paper recursion formulae are proposed for ${\cal N}=2$
$SU(3)$ gauge theory instanton partition function in $\Omega$-background
(Nekrasov's partition function) with $0\le N_f\le 6$
fundamental hypermultiplets as well as for the case with an adjoint
hypermultiplet (${\cal N}=2^*$ theory). As a byproduct all instanton
exact formula is conjectured for the partition function in an
one-parameter family of vacua, which is a natural generalization of
the special vacuum introduced in \cite{Argyres:1999ty} and recently investigated
in \cite{Ashok:2015cba}. The IR-UV relation discovered in
\cite{Billo:2012st,Ashok:2015cba}
was very helpful in finding these results.

Using AGT relation the analogs of Zamolodchikov's
recurrence relations are proposed for the (special) ${\cal W}_3$ $4$-point
blocks on sphere and for the torus $1$-point block. Though CFT point
of view makes many of the features of the recurrence relations natural,
unfortunately rigorous derivations are still lacking.

The organization of the paper is as follows:

In chapter \ref{chapter1}. After a short review of instanton counting in the
theory with $6$ fundamentals, it is shown how investigation of
the poles and residues of the partition function incorporated with
the known UV - IR relation and the insight coming from the 2d CFT
experience leads to the recurrence relation. Then, subsequently
decoupling the hypermultiplets by sending their masses to infinity
corresponding recurrence relations for smaller number of flavours
are found. The simplest case of pure theory ($N_f=0$) is presented
in more details.

Then a similar analysis is carried out and as a result,
corresponding recurrence relation is found for the $SU(3)$,
${\cal N}=2^*$ theory.

In chapter \ref{chapter2}. Using AGT relation, the recurrence relations
are constructed for the $4$-point ${\cal W}_3$ sphere blocks with
two arbitrary and two partially degenerate insertions and for the
torus block with a partially degenerate insertion. In both cases
exact formulae for the large ${\cal W}_3$ current zero mode limit are
presented. It is argued that the location of the poles as well
as the structure of the residues which were instrumental
in finding the recurrence relations of chapter \ref{chapter1}., are related
to the degeneracy condition and the structure of OPE of ${\cal W}_3$ CFT.

\section{Instanton partition function in $\Omega$ background}
\label{chapter1}
\subsection{$SU(3)$ theory with $N_f=6$ fundamental hypermultiplets}
\begin{figure}[t]
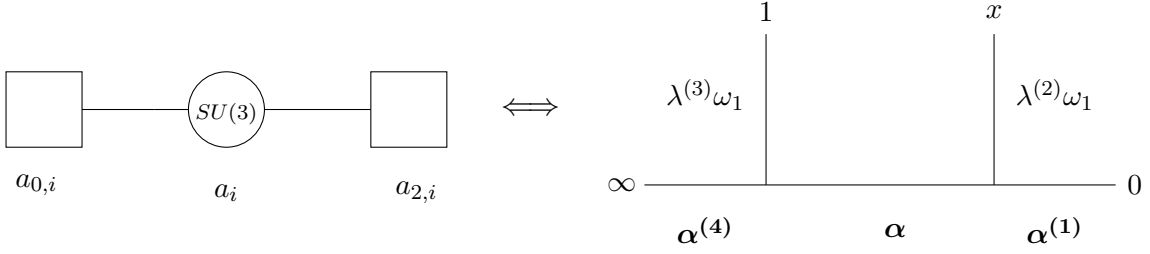

\begin{pgfpicture}{0cm}{0cm}{15cm}{4cm}

\pgfcircle[stroke]{\pgfpoint{3cm}{2.8cm}}{0.5cm}
\pgfputat{\pgfxy(3,2.75)}{\pgfbox[center,center]{\scriptsize{$SU(3)$}}}
\pgfputat{\pgfxy(3,1.7)}{\pgfbox[center,center]{\small{$a_i$}}}
{\color{black}\pgfrect[stroke]{\pgfpoint{0.1cm}{2.3 cm}}{\pgfpoint{1cm}{1cm}}}
\pgfputat{\pgfxy(0.5,1.8)}{\pgfbox[center,center]{\small{$a_{0,i}$}}}
{\color{black}\pgfrect[stroke]{\pgfpoint{4.9cm}{2.3cm}}{\pgfpoint{1cm}{1cm}}}
\pgfputat{\pgfxy(5.5,1.7)}{\pgfbox[center,center]{\small{$a_{2,i}$}}}
\pgfline{\pgfxy(1.1,2.8)}{\pgfxy(2.05,2.8)}
\pgfline{\pgfxy(2.05,2.8)}{\pgfxy(2.5,2.8)}
\pgfline{\pgfxy(3.5,2.8)}{\pgfxy(4.9,2.8)}
\pgfputat{\pgfxy(7,2.8)}{\pgfbox[center,center]{$\Longleftrightarrow$}}
\pgfline{\pgfxy(8.5,1.8)}{\pgfxy(10.1,1.8)}
\pgfline{\pgfxy(10.1,1.8)}{\pgfxy(10.1,3.8)}
\pgfline{\pgfxy(10.1,1.8)}{\pgfxy(13.1,1.8)}
\pgfline{\pgfxy(13.1,1.8)}{\pgfxy(13.1,3.8)}
\pgfline{\pgfxy(13.1,1.8)}{\pgfxy(14.7,1.8)}
\pgfputat{\pgfxy(11.8,1.2)}{\pgfbox[center,center]{\small{$\boldsymbol{\alpha} $}}}
\pgfputat{\pgfxy(9.3,3)}{\pgfbox[center,center]{\small{$\lambda^{(3)}\omega_1$}}}
\pgfputat{\pgfxy(9.3,1.2)}{\pgfbox[center,center]{\small{$\boldsymbol{\alpha^{(4)}} $}}}
\pgfputat{\pgfxy(13.9,3)}{\pgfbox[center,center]{\small{$\lambda^{(2)}\omega_1$}}}
\pgfputat{\pgfxy(13.9,1.2)}{\pgfbox[center,center]{\small{$\boldsymbol{\alpha^{(1)}} $}}}
\pgfputat{\pgfxy(8.2,1.8)}{\pgfbox[center,center]{\small{$\infty$}}}
\pgfputat{\pgfxy(10.1,4.1)}{\pgfbox[center,center]{\small{$1$}}}
\pgfputat{\pgfxy(13.1,4.1)}{\pgfbox[center,center]{\small{$x$}}}
\pgfputat{\pgfxy(14.95,1.8)}{\pgfbox[center,center]{\small{$0$}}}
\pgfclearendarrow
\end{pgfpicture}
\caption{On the left: the quiver diagram for the conformal $SU(3)$
gauge theory with $6$ fundamental hypermultiplets.
On the right: the dual ${\cal W}_3$ conformal block.}
\label{figAGT}
\end{figure}

Graphically this theory can be depicted as a quiver diagram
on the left side of Fig.\ref{figAGT}.
The parameters $a_{0,i}$, $a_{2,i}$ are related to the hypermultiplet
masses while $a_i$ ($i$ runs over $1,2,3$) are the expectation values
of the vector multiplet. The instanton part of the partition
function is given as a sum over
triple of Young diagrams ${\vec{Y}}=(Y_1,Y_2,Y_3)$
(see \cite{Nekrasov:2002qd, Flume:2002az,Bruzzo:2002xf})
\bea
Z=\sum_{\vec{Y}}Z_{\vec{Y}}x^{|\vec{Y}|},
\label{Zinst}
\eea
where $x$ is the exponentiated coupling (the instanton counting
parameter)$, |\vec{Y}|$ is the total number of boxes of Young diagrams.
The coefficients $Z_{\vec{Y}}$ can be represented as
\bea
Z_{\vec{Y}}=\prod_{i,j=1}^3\frac{Z_{bf}(\emptyset ,a_{i,0}|Y_j,a_j)
Z_{bf}(Y_i,a_i|\emptyset ,a_{2,j})}{Z_{bf}(Y_i ,a_i|Y_j,a_j)}
\label{Z_Y}
\eea
where
\bea
Z_{bf}(\lambda,a|\mu,b)&=&\nonumber\\
\prod_{s\in \lambda} (a-b-L_\mu(s)\epsilon_1
+(1+A_\lambda (s))\epsilon_2)\prod_{s\in \mu}
(a-b&+&(1+L_\lambda(s))\epsilon_1-A_\mu (s))\epsilon_2)\,.\qquad\qquad
\label{Z_bf}
\eea
Here $A_\lambda(s)$ ($L_\lambda(s)$) is the distance in vertical
(horizontal) direction from the upper (right)
border of the box $s$ to the outer boundary of the diagram $\lambda$
as demonstrated in Fig. \ref{YD}.
As usual $\epsilon_1$ and $\epsilon_2$ denote the parameters of the $\Omega$ background.
\newcount\tableauRow
\newcount\tableauCol
\def\tableauDim{0.4}
\newenvironment{Tableau}[1]{%
  \tikzpicture[scale=0.7,draw/.append style={loosely dotted,gray},
                      baseline=(current bounding box.center)]
    \tableauRow=-1.5
    \foreach \Row in {#1} {
       \tableauCol=0.5
       \foreach\k in \Row {
         \draw[thin](\the\tableauCol,\the\tableauRow)rectangle++(1,1);
         \draw[black,ultra thick](\the\tableauCol,\the\tableauRow)+(0.5,0.5)node{$\k$};
         \global\advance\tableauCol by 1
       }
       \global\advance\tableauRow by -1
    }
}{\endtikzpicture}
\newcommand\tableau[1]{\begin{Tableau}{#1}\end{Tableau}}
\begin{figure}
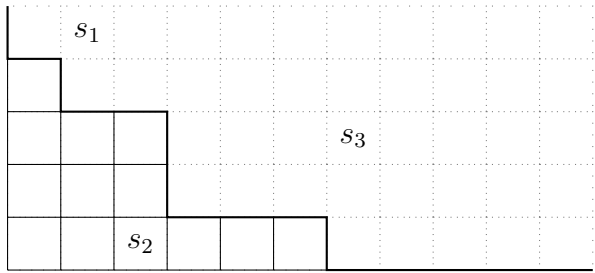

\center
\begin{tabular}{l@{\qquad}l@{\qquad}l}
\begin{Tableau}{{,s_1,,,,,,,,,},{,,,,,,,,,,},{,,,,,,s_3,,,,},
  {,,,,,,,,,,},{,,s_2,,,,,,,,}}
    \draw[thick,solid ,color=black](11,-5)--(6,-5)
    --(6,-4)--(3,-4)--(3,-2)--(1,-2)--(1,-1)--(0,-1)--(0,0);
     \draw[thin,solid ,color=black](0,-1)--(0,-5);
     \draw[thin,solid ,color=black](1,-2)--(1,-5);
     \draw[thin,solid ,color=black](2,-2)--(2,-5);
     \draw[thin,solid ,color=black](3,-4)--(3,-5);
     \draw[thin,solid ,color=black](4,-4)--(4,-5);
     \draw[thin,solid ,color=black](5,-4)--(5,-5);
     \draw[thin,solid ,color=black](0,-5)--(6,-5);
     \draw[thin,solid ,color=black](0,-4)--(3,-4);
     \draw[thin,solid ,color=black](0,-3)--(3,-3);
     \draw[thin,solid ,color=black](0,-2)--(1,-2);
\end{Tableau}
\end{tabular}
\caption{Arm and leg length with respect to the
Young diagram with column lengths $\{4,3,3,1,1,1\}$.
The thick solid line outlines its outer border. $A(s_1)=-2$,
$L(s_1)=-2$, $A(s_2)=2$, $L(s_2)=3$, $A(s_3)=-3$, $L(s_3)=-4$.}
\label{YD}
\end{figure}
Without loss of generality one may assume that $a_1+a_2+a_3=0$. Then
this parameters can be reexpressed in terms of the independent differences
$a_{12}\equiv a_1-a_2$ and $a_{23}\equiv a_2-a_3$
\bea
(a_1,a_2,a_3)=\left(\frac{2a_{12}+a_{23}}{3}\CommaBin -\frac{a_{12}-
a_{23}}{3}\CommaBin -\frac{a_{12}+2a_{23}}{3}\right)\,.
\eea
The masses of $6$ fundamental hypermultiplets can be identified as
\bea
&&m_i=-a_{0,i} \qquad \qquad \quad \,\text{for} \qquad i=1,2,3\,,\nonumber\\
&&m_i=\epsilon_1+\epsilon_2-a_{0,i-3} \quad \text{for}\qquad i=4,5,6\,.
\eea
The advantage of the definition above is that the partition function is symmetric
with respect to permutations of $N_f=6$ masses $m_1,\ldots ,m_6$. For later convenience let us
introduce also notations (elementary symmetric functions of masses)
\bea
T_n=\sum_{1\le i_1<i_2<\cdots <i_n\le N_f}m_{i_1}\cdots m_{i_n}\,.
\eea
Let us fix an instanton number $k$ and perform partial summation in
(\ref{Zinst}) over all diagrams with total number of boxes equal to $k$.
Many spurious poles present in individual terms cancel and one gets a
rational expression whose denominator is
\bea
(\epsilon_1\epsilon_2)^k\prod \left(a_{12}^2-\epsilon_{r,s}^2\right)
 \left(a_{23}^2-\epsilon_{r,s}^2\right) \left((a_{12}+a_{23})^2-\epsilon_{r,s}^2\right),
 \label{denom}
\eea
where the product is over the positive integers $r\ge1$, $s\ge 1$ such
that $rs\le k$ and
\bea
\epsilon_{r,s}=r \epsilon_1+s\epsilon_2\,.
\label{epsrs}
\eea
It is not difficult to check this statement explicitly for small $k$.
Under AGT map this is equivalent to the well known fact that the 2d CFT
blocks as a function of the parameters of the intermediate state acquire
poles exactly at the degeneration points. Anticipating this relation
let us introduce parameters
\bea
&&u=a_{12}^2+a_{12}a_{23}+a_{23}^2,\nonumber\\
&&v=(a_{12}-a_{23})(2a_{12}+a_{23})(a_{12}+2a_{23}).
\label{uvrule}
\eea
We'll see in section \ref{Toda_prel} that $u$ is closely related to
the dimension and $v$ to the {\cal W} zero mode eigenvalue of the
intermediate state.
For what follows it will be crucial to note that the factors of
(\ref{denom}) in terms of newly introduced parameters can be rewritten as
\bea
-27\left(a_{12}^2-\epsilon_{r,s}^2\right)
 \left(a_{23}^2-\epsilon_{r,s}^2\right) \left((a_{12}+a_{23})^2
-\epsilon_{r,s}^2\right)
 =v^2-v_{r,s}(u)^2,
\label{vu_a}
\eea
where
\bea
v_{r,s}(u)=\left(3 \epsilon_{r,s}^2-u\right)\sqrt{4u-3 \epsilon_{r,s}^2}\,\,.
\label{vrs}
\eea
Using (\ref{uvrule}) also in the numerator we can expel the parameters
$a_{12}$, $a_{23}$ in favor of $v$ and $u$. Moreover for fixed $u$
one gets a polynomial dependence on $v$. Thus, to recover the partition
function one needs
\begin{itemize}
\item{the residues at $v=v_{r,s}(u)$};
\item{the asymptotic behaviour of the partition function for a fixed
value of $u$ and large $v$}.
\end{itemize}
\subsubsection{The residues}
It follows from the remarkable identity (\ref{vu_a}) that the
residues at $v=\pm v_{r,s}$ is related to the residue with
respect to the variable $a_{12}$ at $a_{12}=\epsilon_{r,s}$ in a simple
way\footnote{This is a choice of branch of the inverse map
$(v,u)\rightarrow (a_{12},a_{23}$). We could
consider the poles at $a_{23}=\epsilon_{r,s}$ or $a_{12}+a_{23}=\epsilon_{r,s}$
instead.}:
\bea
Res|_{v=\pm v_{r,s}}=\frac{27 \epsilon_{r,s}}{\epsilon_{r,s}+2a_{23}}
\,Res|_{a_{12}=\epsilon_{r,s}}.
\label{res_va}
\eea
To restore the $u$-dependence in right hand side of (\ref{res_va})
due to (\ref{vu_a}) one should substitute
\bea
a_{23}=\frac{-\epsilon_{r,s}\pm \sqrt{4u-3\epsilon_{r,s}^2}}{2}\,.
\label{a23_u}
\eea
A careful examination shows that 
the residue of $k=r s$ instanton term at 
$a_{12}=\epsilon_{r,s}$
receives a nonzero contribution only from the triple $(Y_1,\emptyset ,\emptyset )$, where $Y_1$ 
is a rectangular diagram of size $r\times s$. Using eqs. (\ref{Zinst}), (\ref{Z_Y}), 
(\ref{Z_bf}) it is straightforward 
to evaluate this contribution. The result 
has a nice factorized form
\bea
Res|_{a_{12}=\epsilon_{r,s}}\,Z_{r\cdot s}&=&
-\prod_{i=1-r}^r\sideset{}{'}\prod_{j=1-s}^s\epsilon_{i,j}^{-1}\nonumber\\
&\times &\prod_{i=1}^{r}\prod_{j=1}^{s}
\frac{\prod_{f=1}^{6}
\left(m_f+\frac{1}{3}\,a_{23}+\frac{2}{3}\,\epsilon_{r,s}
-\,\epsilon_{i,j}\right)}
{(a_{23}+\epsilon_{r-i,s-j})(a_{23}+\epsilon_{i,j})}\,,
\label{res_a}
\eea
where the prime over the product means that the term
with $i=j=0$ should be omitted\footnote{
In generic $SU(n)$ case with no hypers a nice formula has
been found earlier \cite{Morales:unpablished} for the
{\it multiple} residues at the values of parameters
$a_{1,n},a_{2,n}\ldots a_{n-1,n}$ specialized
as $a_{i,j}=\epsilon_{r_i,s_j}$.
Unfortunately, these residues
alone are not sufficient to derive a recurrence relation for
the partition function.
}.
\subsubsection{Large $v$ limit}
Now let us consider the limit $v\rightarrow \infty $ for fixed
$u$. This is equivalent to choosing
\bea
a_{23}=\frac{\sqrt{4u-3a_{12}}-a^2_{12}}{2}
\label{special_a}
\eea
and taking large $a_{12}$ limit. Here are the first few terms of
this expansion
\bea
a_{23}=e^{-\frac{i\pi}{3}} a_{12}
-\frac{i u}{\sqrt{3} a_{12}}
-\frac{i u^2}{3 \sqrt{3} a_{12}^3}-\frac{2 i u^3}{9 \sqrt{3} a_{12}^5}
-\frac{5 i u^4}{27 \sqrt{3} a_{12}^7}
-\frac{14 i u^5}{81 \sqrt{3} a_{12}^9}+\cdots\quad
\label{special_a_exp}
\eea
I performed instanton calculation in this limit up to the order $x^5$.
The result up to the order $x^4$ reads:
\bea
&&\epsilon_1\epsilon_2\log Z\sim\nonumber\\
&&x \left(\frac{m_1 \epsilon}{3}-\frac{m_2}{3}-\frac{2 \epsilon^2}{9}-\frac{4 u}{27}\right)
+x^2 \left(\frac{5 m_1 \epsilon}{27}-\frac{m_1^2}{54}
-\frac{7 m_2}{54}-\frac{10 \epsilon^2}{81}-\frac{14 u}{243}\right)\qquad\nonumber\\
&&+x^3 \left(\frac{283 m_1 \epsilon}{2187}-\frac{40 m_1^2}{2187}-\frac{163 m_2}{2187}-\frac{566 \epsilon^2}{6561}-\frac{1948 u}{59049}\right)\nonumber\\
&&+x^4 \left(\frac{655 m_1 \epsilon}{6561}-\frac{433 m_1^2}{26244}-\frac{1321 m_2}{26244}-\frac{1310 \epsilon^2}{19683}
-\frac{3931 u}{177147}\right)+\cdots\, ,
\label{Z_asymp}
\eea
where (and further on) for shortness I use the notation
$\epsilon=\epsilon_1+\epsilon_2$.
Notice that at $u=0$ the choice of VEV (\ref{special_a}), (\ref{special_a_exp})
coincides with the special vacuum investigated in
\cite{Argyres:1999ty,Ashok:2015cba}. In \cite{Billo:2012st,Ashok:2015cba}
an exact relation between the UV coupling and effective IR coupling has
been established. It was shown that a central role is played by the
congruence subgroup $\Gamma_1(3)$ of the duality group $SL(2,\mathbb{Z})$
\cite{koblitz2012introduction,apostol2012modular}
and that the relation
\bea
x=-27 \left(\frac{\eta(q^3)}{\eta(q)}\right)^{12}
\label{x_q}
\eea
between $x=\exp 2\pi i\tau_{uv}$ and
$q=\exp 2\pi i\tau_{ir}$, where
$\eta(q)$ is Dedekind's eta function
\bea
\eta(q)=q^{1\over 24}\prod_{n=1}^\infty (1-q^n)
\eea
is valid. It should not come as a surprise also that the unique
degree $1$ modular form of $\Gamma_1(3)$
\bea
f_1(q)=\left(\left(\frac{\eta^3(q)}{\eta(q^3)}\right)^3
+27 \left(\frac{\eta^3(q^3)}{\eta(q)}\right)^3\right)^{1/3}
\eea
and its "ingredients" have a role to play. Indeed the expression
\bea
\epsilon_1\epsilon_2\log \left(\left(-\frac{x}{27q}\right)^
{\frac{u}{3\epsilon_1\epsilon_2}}
\left(\frac{\eta(q^3)}{\eta^3(q)}\right)^{\frac{3 T_2-T_1^2}{\epsilon_1\epsilon_2}}
f_1(q)^{\frac{T_1^2-3T_1\epsilon+2\epsilon^2}{2\epsilon_1\epsilon_2}}
\right)
\label{Z_asymp_exact}
\eea
nicely matches the expansion (\ref{Z_asymp}) up to quite high orders
in $q$ and there is little doubt that the argument of logarithm in
(\ref{Z_asymp_exact}) indeed gives the large
$v$ limit of the partition function exactly.
\subsubsection{The recurrence relation}
Using AGT relation it is not difficult to establish that the residue
of the partition function at $v=\pm v_{r,s}(u)$ is proportional to
the partition function with expectation values specified  as
\bea
v\rightarrow \pm v_{r,-s}(u-3\epsilon_1\epsilon_2\,rs)
\,;\qquad u\rightarrow u-3\epsilon_1\epsilon_2\,rs \,.
\eea
On CFT side these are exactly the values corresponding to the null vector
built from the given degenerate intermediate state related to the choice
$v=\pm v_{r,s}(u)$.
Let us represent the partition function as
\bea
Z(v,u,q)=\left(-\frac{x}{27q}\right)^
{\frac{u}{3\epsilon_1\epsilon_2}}
\left(\frac{\eta(q^3)}{\eta^3(q)}\right)^{\frac{3 T_2-T_1^2}{\epsilon_1\epsilon_2}}
f_1(q)^{\frac{T_1^2-3T_1\epsilon+2\epsilon^2}{2\epsilon_1\epsilon_2}}
H(v,u|q).
\label{Z_H}
\eea
Note that
\bea
H(v,u|q)=1+O(v^{-1}).
\eea
Incorporating information about residues establish above we
finally arrive at the recurrent relation
\bea
H(v,u|q)=1+\sum_{r,s=1}^\infty\sum_{\sigma=\pm}\frac{(-27q)^{rs}
R^{(\sigma)}_{r,s}(u)}{v-\sigma v_{r,s}(u)}
\,H\left(\sigma v_{r,-s}(u-3\epsilon_1\epsilon_2\,rs),
u-3\epsilon_1\epsilon_2\,rs\right |q),\nonumber\\
\label{recursion}
\eea
where due to eqs. (\ref{res_va}), (\ref{res_a})
\bea
R_{r,s}^{(\pm)}&=&\frac{ 27\epsilon_{r,s}\left(u-\epsilon_{r,s}^2\right)}
{\mp \sqrt{4u-3\epsilon_{r,s}^2}}
\prod_{i=1-r}^r\sideset{}{'}\prod_{j=1-s}^s\epsilon_{i,j}^{-1}\nonumber\\
&\times &\prod_{i=1}^{r}\prod_{j=1}^{s}\frac{\prod_{l=1}^{N_f}
\left(m_l-\frac{1}{2}\,\epsilon_{2i-r,2j-s}\pm\frac{1}{6}\,
\sqrt{4u-3\epsilon_{r,s}^2}\right)}
{u-\epsilon_{r,s}^2+\epsilon_{i,j}\epsilon_{r-i,s-j}}\,.
\label{res_v}
\eea
Using the recurrence relation I have computed the partition function
up to the order $x^8$ and compared it with the result
of the direct instanton calculation. The agreement was perfect.
\subsection{$N_f<6$ cases}
\label{nf<6}
It is straightforward to decouple some of $6$ hypermultiplets
sending their masses to infinity.

Let us choose $m_{N_f+1}=\cdots=m_{6}=\Lambda$, renormalize
the coupling constant as $x\rightarrow -\frac{x}{\Lambda^{6-N_f}}$
and take the large
$\Lambda$ limit\footnote{The minus sign is due to a subtle
difference between fundamental and anti-fundamental hypermultiplets.
With this sign included we get $N_f$ anti-fundamentals in conventions
of \cite{Alday:2009aq}.}. The net effect is that instead of the
recursion relation (\ref{recursion}) one obtains
\bea
H(v,u|x)=1+\sum_{r,s=1}^\infty\sum_{\sigma=\pm}\frac{(-x)^{rs}
R^{(\sigma)}_{r,s}(u)}{v-\sigma v_{r,s}(u)}
\,H\left(\sigma v_{r,-s}(u-3\epsilon_1\epsilon_2\,rs),
u-3\epsilon_1\epsilon_2\,rs\right |x),\nonumber\\
\label{recursion_less}
\eea
where for the residues the same formula
(\ref{res_v}) with appropriate number of hypermultiplets $N_f$ is valid.
The relation between $Z$ and $H$ becomes much simpler.
Using eq. (\ref{Z_asymp}) we
immediately see that for $N_f=5$ the appropriate relation is
\bea
Z_{N_f=5}=\exp \left(\frac{x\,(18(T_1-\epsilon)-x)}
{54\epsilon_1\epsilon_2}\right)H(v,u|x),
\eea
and, for $N_f=4$:
\bea
Z_{N_f=4}=\exp \left(\frac{x}{3\epsilon_1\epsilon_2}\right)\,H(v,u|x).
\eea
Finally in the cases $N_f=0,1,2,3$ the functions $Z$ and $H$ simply coincide.
\subsubsection{Pure $SU(3)$ theory}
\label{Nf=0}
This is the simplest case. It is easy to realize that the
partition function is even with respect to the parameter $v$,
so that the expansion (\ref{recursion}) can be organized
according to the poles in the variable $v^2$:
\bea
Z(v^2,u|x)=1+\sum_{r,s=1}^\infty\frac{(-x)^{rs}
R_{r,s}(u)}
{v^2-v^2_{r,s}(u)}
\,Z\left(v^2_{r,-s}(u-3\epsilon_1\epsilon_2\,rs),
u-3\epsilon_1\epsilon_2\,rs\right |x),\qquad
\label{recursion_pure}
\eea
where
\bea
R_{r,s}=54\epsilon_{r,s}\left(u-\epsilon_{r,s}^2\right)
\left(u-3\epsilon_{r,s}^2\right)
\prod_{i=1-r}^r\sideset{}{'}\prod_{j=1-s}^s\epsilon_{i,j}^{-1}
\prod_{i=1}^{r}\prod_{j=1}^{s}\left(
u-\epsilon_{r,s}^2+\epsilon_{i,j}\epsilon_{r-i,s-j}
\right)^{-1}.\nonumber\\
\label{res_v2}
\eea
\subsection{$N=2^*$ theory}
\label{rec2*}
The analysis of the $SU(3)$ theory with an adjoint hypermultiplet
can be carried out in a similar manner. The coefficients $Z_{\vec{Y}}$
of the instanton partition function (\ref{Zinst})
in this case is given by
\bea
Z_{\vec{Y}}=\prod_{i,j=1}^3\frac{Z_{bf}(Y_i ,a_i-m|Y_j,a_j)}
{Z_{bf}(Y_i ,a_i|Y_j,a_j)}\,,
\label{Z_inst_adj}
\eea
where $m$ is the mass of the adjoint hypermultiplet. The
structure of poles is the same as in the previous cases.
Due to symmetry under permutation
$a_{12}\leftrightarrow a_{23}$ the partition function, as
in the case of pure theory, is a function of $v^2$.
The residue of the $k=rs$ instanton charge sector of the
partition function at $v^2=v^2_{r,s}$ and fixed $u$ is
related to the residue in variable $a_{12}$ at
$a_{12}=\epsilon_{r,s}$ (with $a_{23}$ fixed)
\bea
Res|_{v^2=v^2_{r,s}}=-54\,\epsilon_{r,s}
(a_{23}^2-\epsilon^2_{r,s})(2\epsilon_{r,s}a_{23}+a_{23}^2)
\,Res|_{a_{12}=\epsilon_{r,s}}.
\label{res_va_adj}
\eea
As in the case of fundamental hypermultiplets  
the residue of $k=r s$ instanton term at $a_{12}=\epsilon_{r,s}$ 
receives a nonzero contribution only from the triple of Young diagrams 
$(Y_1,\emptyset ,\emptyset )$ with $Y_1$ 
being a rectangular diagram of size $r\times s$. A direct calculation, using eqs. 
(\ref{Z_bf}), (\ref{Z_inst_adj})  shows that
\bea
Res|_{a_{12}=\epsilon_{r,s}}\,Z_{r\cdot s}&=&
\prod_{i=1-r}^r\sideset{}{'}\prod_{j=1-s}^s
\frac{\epsilon_{i,j}-m}{\epsilon_{i,j}}\nonumber\\
&\times &\prod_{i=1}^{r}\prod_{j=1}^{s}
\frac{(a_{23}+\epsilon_{r-i,s-j}+m)(a_{23}+\epsilon_{i,j}-m)}
{(a_{23}+\epsilon_{r-i,s-j})(a_{23}+\epsilon_{i,j})}\,.
\label{res_a_adj}
\eea
Investigation of the large $v^2$ behavior in this case is simpler
compared to the theory with $6$ fundamentals. Computations in first
few instanton orders shows that (in this section a more
conventional notation $q$ instead of $x$ for the instanton
counting parameter is restored)
\bea
\epsilon_1\epsilon_2 \log Z_{{\cal N}=2^*}= -3(m-\epsilon_1)(m-\epsilon_2)
\log \left(q^{-\frac{1}{24}}\,\eta(q)\right)+O(v^{-2}).
\label{Z_asymp_adj}
\eea
This is a suggestive result. Recall that in the case of $SU(2)$
gauge group one gets the same answer with the only difference
that the overall factor $3$ is replaced by $2$ \cite{Poghossian:2009mk}.

Further steps are straightforward. Introducing the function $H$
via
\bea
Z_{N=2^*}=\left(q^{-\frac{1}{24}}\,\eta(q)\right)^{-\frac{3(m-\epsilon_1)
(m-\epsilon_2)}{\epsilon_1\epsilon_2}}\,H(v^2,u,q)
\label{ZH_adj}
\eea
we get the recurrence relation
\bea
H(v^2,u|q)=1+\sum_{r,s=1}^\infty\frac{q^{rs}
R_{r,s}(u)}
{v^2-v^2_{r,s}(u)}
\,H\left(v^2_{r,-s}(u-3\epsilon_1\epsilon_2\,rs),
u-3\epsilon_1\epsilon_2\,rs\right |q),\qquad
\label{recursion_adj}
\eea
where
\bea
R_{r,s}&=&-54\,m \epsilon_{r,s}\left(u-\epsilon_{r,s}^2\right)
\left(u-3\epsilon_{r,s}^2\right)
\prod_{i=1-r}^r\sideset{}{'}\prod_{j=1-s}^s
\frac{\epsilon_{i,j}-m}{\epsilon_{i,j}}\nonumber\\
&\times &\prod_{i=1}^{r}\prod_{j=1}^{s}
\frac{u-\epsilon^2_{r,s}+(\epsilon_{i,j}-m)(\epsilon_{r-i,s-j}+m)}
{u-\epsilon^2_{r,s}+\epsilon_{i,j}\epsilon_{r-i,s-j}}\,.
\label{res_v2_adj}
\eea
This recurrence relation has been checked by
instanton calculation up to the order $q^{10}$.

\section{Recurrence relation for ${\cal W}_3$ conformal blocks}
\label{chapter2}
In this section using AGT relations
\cite{Alday:2009aq,Wyllard:2009hg,Fateev:2011hq}
 the recurrence relations for
${\cal N}=2$ SYM partition functions will be translated into
recurrence relations for certain ${\cal W}_3$-algebra four-point conformal
blocks on sphere (AGT counterpart of $N_f=6$ theory) and one-point
torus blocks (AGT dual of ${\cal N}=2^*$). This recurrence relations
generalize Alexei Zamolodchikov's famous relation established for the
four point Virasoro conformal blocks
\cite{Zamolodchikov:1985ie,Zamolodchikov:1987tmf}. The recurrent relation
for Virasoro $1$-point torus block was proposed in
\cite{Poghossian:2009mk} (see also \cite{Hadasz:2009db}). It should be
emphasised nevertheless, that the ${\cal W}_3$ blocks considered here are
not quite general, two of four primary fields of the sphere block
as well as that of the $1$-point torus block are specific. The charge vectors defining their dimensions and ${\cal W}_3$ zero-mode eigenvalues
are taken to be multiples of the highest weight of the fundamental
(or anti-fundamental) representation of $SU(3)$. Unfortunately
effective methods to understand generic ${\cal W}$-blocks (to my knowledge)
are still lacking.
\subsection{Preliminaries on $A_2$ Toda CFT}
\label{Toda_prel}
These are 2d CFT theories which, besides the spin $2$ holomorphic
energy momentum current ${\cal W}^{(2)}(z)\equiv T(z)$ are endowed with additional higher spin
$s=3$ current ${\cal W}^{(3)}$
\cite{Zamolodchikov:1985wn,Fateev:1987zh,Bilal:1988ze}. The Virasoro
central charge is conventionally parameterised as
\[
c=2+24 Q^2\,,
\]
where the "background charge" $Q$ is given by
\[
Q=b+\frac{1}{b}\,,
\]
and $b$ is the
dimensionless coupling constant of Toda theory. In what follows it would
be convenient to represent roots, weights and Cartan elements of the
Lie algebra $A_{2}$ as $3$-component vectors satisfying the condition that
the sum of the components is zero. It is assumed also that the scalar product is
the usual Kronecker one. Obviously this is equivalent
to a more conventional representation of these quantities as diagonal traceless
$3\times 3$ matrices with pairing given by trace. In this representation
the Weyl vector is given by
\bea
\boldsymbol{\rho}=\left(1,0,-1\right).
\eea

For further reference let us quote here explicit expressions for the highest weight $\boldsymbol{\omega}_1$ of the first fundamental representation and for its complete set of weights $\mathbf{h}_1, \mathbf{h}_2,\mathbf{h}_3$
\bea
&&\boldsymbol{\omega}_1=\left(\frac{2}{3}\CommaBin -\frac{1}{3}\CommaBin
-\frac{1}{3}\right),\nonumber\\
&&(\mathbf{h}_l)_i=\delta_{l,i}-1/3\,.
\eea
The primary fields $V_{\boldsymbol{\alpha}}$ (in this paper we concentrate only on
the left moving holomorphic parts) are parameterized by vectors $\boldsymbol{\alpha}$ with
vanishing center of mass. Their conformal wights are given by
\bea
h_{\boldsymbol{\alpha}}=\frac{(\boldsymbol{\alpha} ,2Q\boldsymbol{\rho}
-\boldsymbol{\alpha})}{2}\,.
\eea
Sometimes it is convenient to parameterize primary fields (or states)
in terms of the Toda momentum vector $\mathbf{p}=Q\boldsymbol{\rho}
-\boldsymbol{\alpha}$ instead of $\boldsymbol{\alpha}$.
In what follows a special role is played by the fields
$V_{\lambda \boldsymbol{\omega}_1}$
with dimensions
\bea
h_{\lambda \boldsymbol{\omega}_1}=\lambda \left(Q-\frac{\lambda}{3}\right)\,.
\label{dim_lambda}
\eea
For generic $\lambda$ these fields admit a single null vector at the first level.

Besides the dimension, the fields are characterized also by the zero mode eigenvalue
of the ${\cal W}_3$ current
\bea
w=-\frac{i}{27} \sqrt{\frac{48}{22+5c}}\,\,v\,,
\eea
where $v$ is defined in terms of the momentum vector $\mathbf{p}$ as
\bea
v=27 p_1p_2p_3=(p_{12}-p_{23})(p_{12}+2p_{23})(2p_{12}+p_{23})
\label{v_p}
\eea
and $p_{12}=p_1-p_2$, $p_{23}=p_2-p_3$.
It is convenient to introduce also the parameter
\bea
u=p_{12}^2+p_{23}^2+p_{12}p_{23}
\label{u_p}
\eea
so that the conformal dimension (\ref{dim_lambda}) can be rewritten as
\bea
h=Q^2-\frac{u}{3}\,.
\eea
The pair $v,u$ characterizes primary fields more faithfully, than the charge
vector, since they are invariant under the Weyl group action.
\subsubsection{Sphere $4$-point block}
The object of our interest in this section will be the conformal block
\bea
\langle V_{\boldsymbol{\alpha}_4}(\infty)
V_{\lambda_3 \boldsymbol{\omega}_1}(1)V_{\lambda_2\boldsymbol{\omega}_1}(x)
V_{\boldsymbol{\alpha}_1}(0) \rangle_\mathbf{p} \sim x^{h_{\boldsymbol{\alpha}}-h_1-h_2}
G(v,u|x)\,,
\eea
where $\langle \cdots\rangle_{\mathbf{p}} $ denotes the holomorphic
part of the
correlation function with a specified intermediate state of
momentum $\mathbf{p}=Q\boldsymbol{\rho} -\boldsymbol{\alpha} $.
It is assumed that the function $G(v,u|x)$ is normalized so that
$G(v,u|x)=1+O(x)$ (we explicitly
display only dependence on the parameters $v,u$, which specify the
intermediate state).
Due to AGT relation, the function $G(v,u|x)$ is directly connected
to the instanton partition function of $SU(3)$ gauge theory with
$N_f=6$ hypermultiplets discussed earlier (see Fig.\ref{figAGT}).
Here is the
map between parameters of the CFT and Gauge Theory (GT) sides:
\bea
b&=&\sqrt{\frac{\epsilon_1}{\epsilon_2}};\qquad u_{CFT}=
\frac{u_{GT}}{\epsilon_1\epsilon_2};\qquad
v_{CFT}=\frac{v_{GT}}{(\epsilon_1\epsilon_2)^{3/2}};\qquad \\
\label{lambda_m}
\lambda^{(2)}&=&\frac{3 \epsilon-m_4-m_5-m_6}{\sqrt{\epsilon_1\epsilon_2}};\qquad
\lambda^{(3)}=\frac{m_1+m_2+m_3}{\sqrt{\epsilon_1\epsilon_2}};\\
\label{p_m1}
\mathbf{p}^{(1)}&=&Q\boldsymbol{\rho}-\boldsymbol{\alpha}^{(1)}\nonumber\\
\qquad &=&\left(
\frac{-2m_4+m_5+m_6}{\sqrt{\epsilon_1\epsilon_2}},
\frac{-2m_5+m_4+m_6}{\sqrt{\epsilon_1\epsilon_2}},
\frac{-2m_6+m_4+m_5}{\sqrt{\epsilon_1\epsilon_2}}
\right);\\
\label{p_m4}
\mathbf{p}^{(4)}&=&Q\boldsymbol{\rho}-\boldsymbol{\alpha}^{(4)}\nonumber\\
\qquad &=&\left(
\frac{-2m_1+m_2+m_3}{\sqrt{\epsilon_1\epsilon_2}},
\frac{-2m_2+m_1+m_3}{\sqrt{\epsilon_1\epsilon_2}},
\frac{-2m_3+m_1+m_2}{\sqrt{\epsilon_1\epsilon_2}}
\right)\,.
\eea
Under this identification of parameters the relation between the gauge
theory (with $N_f=6$ fundamentals) partition function and the CFT
conformal block is very simple:
\bea
Z=(1-x)^{\lambda^{(3)}\left(Q-\frac{1}{3}\,\lambda^{(2)}\right)}\,G\,.
\label{AGT}
\eea
Now it is quite easy to rephrase the recurrence relation for the
partition function in terms of CFT language. Define a function $H(v,u|q)$
through
\bea
G(v,u|x)=\left(-\frac{x}{27q}\right)^
{\frac{u}{3}}
\left(\frac{\eta(q^3)}{\eta^3(q)}\right)^{3(h_1+h_4)-6Q^2}
f_1(q)^{\frac{-3(h_2+h_3)+2Q^2}{2}}H(v,u|q),
\label{G_H}
\eea
where $q$ and $x$ are related as in (\ref{x_q}). Then,
due to (\ref{Z_H}), (\ref{recursion}), (\ref{AGT}) and (\ref{G_H})
for $H(v,u|q)$ we get essentially the same recurrence relation
(\ref{recursion})
\bea
H(v,u|q)=1+\sum_{r,s=1}^\infty\sum_{\sigma=\pm}\frac{(-27q)^{rs}
R^{(\sigma)}_{r,s}(u)}{v-\sigma v_{r,s}(u)}
\,H\left(\sigma v_{r,-s}(u-3rs),
u-3rs\right |q),\nonumber\\
\label{recursion_CFT}
\eea
where similar to (\ref{vrs})
\bea
v_{r,s}(u)=(3Q_{r,s}^2-u)\sqrt{4u-3Q_{r,s}^2}
\label{vrs_CFT}
\eea
with (cf. (\ref{epsrs}) )
\bea
Q_{r,s}=b r+\frac{s}{b}
\label{Qrs}
\eea
and the residues are given by
\bea
R_{r,s}^{(\pm)}&=&\frac{ 27Q_{r,s}\left(u-Q_{r,s}^2\right)}
{\mp \sqrt{4u-Q_{r,s}^2}}
\prod_{i=1-r}^r\sideset{}{'}\prod_{j=1-s}^sQ_{i,j}^{-1}\nonumber\\
&\times &\prod_{i=1}^{r}\prod_{j=1}^{s}\frac{\prod_{l=1}^{6}
\left(\mu_l-\frac{1}{2}\,Q_{2i-r,2j-s}\pm\frac{1}{6}\,
\sqrt{4u-3Q_{r,s}^2}\right)}
{u-Q_{r,s}^2+Q_{i,j}Q_{r-i,s-j}}\,,
\label{res_v_CFT}
\eea
where CFT counterparts of gauge theory masses $\mu_l=m_l
/\sqrt{\epsilon_1\epsilon_2}$ are related to the parameters of
the inserted fields via (\ref{lambda_m})-(\ref{p_m4}).

It follows from the analog of the Kac determinant for ${\cal W}_3$-algebra
\cite{Watts:1989bn}, that the conformal block truncated up to the
order $x^k$ should have simple poles in the variable $v$
(for $u$ fixed) located at $v=\pm v_{r,s}(u)$ with $r\ge 1$,
$s\ge 1$ and $r\, s\le k$. The relation
\bea
v^2-v^2_{r,s}(u)=0
\eea
among parameters $v$, $u$ is
the condition of existence of a null vector at the level $r s$.
This null vector originates a ${\cal W}_3$-algebra representation with
parameters
\bea
u\rightarrow u-3rs\,;\qquad v\rightarrow \pm v_{r,-s}(u-3 rs).
\label{pole_parameters}
\eea
Though we arrived to the recurrence relation starting from the gauge
theory side, in fact many features of this relation are transparent
from the CFT side and it is reasonable to expect that a rigorous
proof may be found generalizing arguments of Alexei Zamolodchikov
from Virasoro to the ${\cal W}$-algebra case.
Indeed (\ref{recursion_CFT}) states that the residues at the poles
$v=\pm v_{r,s}(u)$ (\ref{vrs_CFT}), are proportional to
the conformal block with internal channel parameters
(\ref{pole_parameters}) corresponding to the null vector at the
level $rs$.

The factor $R_{r,s}^{(\pm)}$
(\ref{res_v_CFT}) also has many expected features. Its
denominator vanishes exactly when the parameter $u$ is specified so
that a second independent degenerate state arises. The factors
in the numerator reflect the structure of OPE with degenerate
field (see \cite{Fateev:2007ab}) exactly as it was in the case of Virasoro
block considered by Alexei Zamolodchikov. It seems more subtle to
justify presence of the $u$ independent factors $Q_{i,j}^{-1}$.

Our result predicts the following large $v$ behavior of the ${\cal W}_3$ block
\bea
G(v,u|x)\sim \left(-\frac{x}{27q}\right)^
{\frac{u}{3}}
\left(\frac{\eta(q^3)}{\eta^3(q)}\right)^{3(h_1+h_4)-6Q^2}
f_1(q)^{\frac{-3(h_2+h_3)+2Q^2}{2}}
+O(v^{-1}).
\label{G_asymp_exact}
\eea
A good starting point to prove this relation might be the deformed
Seiberg-Witten curve DSFT
\cite{Poghossian:2010pn,Fucito:2011pn,Nekrasov:2013xda}
or, equivalently, the quasiclassical null vector decoupling equation
for ${\cal W}$-blocks derived in \cite{Poghossian:2016rzb}.
\subsubsection{Torus $1$-point block}
Since the torus $1$-point block (below $\boldsymbol{\alpha}$ is
the charge parameter of the intermediate states)
\bea
{\cal F}_{\boldsymbol{\alpha}}^{\lambda}(q)=
q^{\frac{c}{24}-h_{\boldsymbol{\alpha}}}\tr_{\boldsymbol{\alpha}}
\left(
q^{L_0-\frac{c}{24}}V_{\lambda \boldsymbol{\omega}_1}(1)\right)
\eea
is related to the partition function of the gauge theory with adjoint
hypermultiplet via \cite{He:2012bi}
\bea
Z_{{\cal N}=2^*}=
\left(q^{-\frac{1}{24}}\,\eta(q)\right)^{-\lambda(Q-\frac{\lambda}{3})
-1}\,{\cal F}_{\boldsymbol{\alpha}}^{\lambda}(q)\,.
\eea
The parameter $\lambda $ is related to the adjoint hypermultiplet mass $m$:
\bea
\lambda=\frac{3m}{\sqrt{\epsilon_1\epsilon_2}}
\eea
and as earlier the intermediate momentum parameter
$\mathbf{p}=Q\boldsymbol{\rho}-\boldsymbol{\alpha}$ is related
to the VEV of the vector multiplet $\mathbf{a}$ as
\bea
p_i=\frac{a_i}{\sqrt{\epsilon_1\epsilon_2}}; \qquad i=1,2,3\,.
\eea
Thus, comparing with (\ref{ZH_adj}), (\ref{recursion_adj}),
(\ref{res_v2_adj}), we see
that the function $H(v^2,u,q)$ defined by
the equality
\bea
{\cal F}_{\boldsymbol{\alpha}}^{\lambda}(q)=
\left(q^{-\frac{1}{24}}\,\eta(q)\right)^{-2}\,H(v^2,u,q)\,,
\label{H_torus}
\eea
($v$ and $u$ in terms of the momentum
$p$ were defined in (\ref{v_p}), (\ref{u_p}))
satisfies the recurrence relation
\bea
H(v^2,u|q)=1+\sum_{r,s=1}^\infty\frac{q^{rs}
R_{r,s}(u)}
{v^2-v^2_{r,s}(u)}
\,H\left(v^2_{r,-s}(u-3rs)\,,
u-3rs\right |q),\qquad
\label{recursion_torus}
\eea
where
\bea
R_{r,s}&=&-18\,\lambda\, Q_{r,s}\left(u-Q_{r,s}^2\right)
\left(u-3Q_{r,s}^2\right)
\prod_{i=1-r}^r\sideset{}{'}\prod_{j=1-s}^s
\frac{Q_{i,j}-\frac{\lambda}{3}}{Q_{i,j}}\nonumber\\
&\times &\prod_{i=1}^{r}\prod_{j=1}^{s}
\frac{u-Q^2_{r,s}+(Q_{i,j}-\frac{\lambda}{3})(Q_{r-i,s-j}+\frac{\lambda}{3})}
{u-Q^2_{r,s}+Q_{i,j}Q_{r-i,s-j}}\,.
\label{res_v2_torus}
\eea
\section{Summary and discussion}
To summarize let me quote the main results of this paper:
\begin{itemize}
\item{the recurrence relation (see
(\ref{Z_asymp_exact}), (\ref{recursion}), (\ref{res_v}))
for the instanton partition function
of ${{\cal N}=2}$ $SU(3)$ gauge theory with $6$ fundamental
hypermultiplets. This recurrence relation suggests an exact in all
instanton orders formula for the partition function and prepotential
for the theory in a generalized version of the special vacuum considered in
\cite{Argyres:1999ty,Ashok:2015cba}};
\item{recurrence relations for smaller number of hypermultiplets
(see section \ref{nf<6}) and for pure $N_f=0$ theory
(section \ref{Nf=0})};
\item{recurrence relations for the theory with an adjoint
hypermultiplet, commonly referred as ${\cal N}=2^*$ theory
(see section \ref{rec2*})};
\item{the analogs of Zamolodchikov's recurrence relations are
constructed for $4$-point sphere ${\cal W}_3$-blocks with two arbitrary
and two partially degenerate insertions (see (\ref{G_H}),
(\ref{recursion_CFT}), (\ref{res_v_CFT}))
and for the torus ${\cal W}_3$-block with a partially degenerate insertion
(see (\ref{H_torus}),(\ref{recursion_torus}), (\ref{res_v2_torus})).
For both cases recursion formulae provide explicit expressions
for the large ${\cal W}_3$ zero mode limit.}
\end{itemize}
Though many details of the recurrence relations are transparent either
from the 4d gauge theory or from the 2d CFT points of view, still full
derivation is lacking. I hope to come back to these questions
in a future publication.

Of course, generalization to the case of generic $SU(n)$/${\cal W}_n$ cases
would be an interesting development.
\vspace{1cm}

\section*{Acknowledgments}

I am grateful to G.~Bonelli, F.~Fucito, F.~Morales, A.~Tanzini  for stimulating
discussions and for hospitality at the university of Rome "Tor Vergata"
and SISSA, Trieste during February of this year, where the initial ideas of this paper
emerged.

This work was partially supported by the Armenian State Committee of Science
in the framework of the research project 15T-1C308.

\providecommand{\href}[2]{#2}\begingroup\raggedright\endgroup


\end{document}